\documentclass[twocolumn,aps,prl,showpacs,amsmath,amssymb,floatfix]{revtex4}
\usepackage{graphics,color}
\begin{document}

\title{Influence of Correlated Hybridization on the Conductance of Molecular Transistors}

\author{J.C. Lin$^{1}$, F.B. Anders$^{2,3}$, and D.L. Cox$^{1}$}

\affiliation{$^{1}$Department of Physics, University of California, Davis, California 95616 \\
$^{2}$Institut f\"ur Theoretische Physik, Universit\"at Bremen, P.O. Box 330 440, D-28334 Bremen, Germany \\
$^{3}$Theoretische Physik, Universit\"{a}t des Saarlands, 66041 Saarbr\"{u}cken, Germany}

\date{\today}   

\begin{abstract}
We study the spin-1/2 single-channel Anderson impurity model with correlated 
(occupancy dependent) hybridization for molecular transistors using the 
numerical renormalization-group method. Correlated hybridization can induce nonuniversal deviations in the normalized zero-bias conductance and, for some parameters, modestly enhance the spin polarization of currents in applied magnetic field. Correlated hybridization can also explain a gate-voltage dependence to the Kondo scale similar to what has been observed in recent experiments. 
\end{abstract}


\maketitle

With the technological advances allowing for the construction of single-molecule
 transistors based upon transition-metal 
complexes \cite{Liang2002,Pasupathy2004,Natelson}, there is considerable 
interest in the out of equilibrium properties of the Anderson model. 
This model is used to describe the coupling of electrons in localized and 
strongly interacting states of an atom or molecule with electrons in weakly 
interacting extended states of the leads or host metal (see the schematic 
of Fig. 1). 
Most interest has focused on the resonant transport regime of the model 
associated with the local moment or Kondo limit of the Anderson model. 
In this case, correlated (occupancy dependent) hybridization has been invoked 
as a potential explanation for anomalies in the conductance \cite{Meir2002}, the unusual gate-voltage dependence of the measured Kondo temperature 
scale \cite{Natelson} {\it not} described by the simplest version of the 
Anderson model, and to suggest a possible conductance enhancement via local 
pairing effects \cite{Guinea2003}. In parallel, it has been demonstrated for transition metal-complexes that the correlated hybridization can be very large, comparable to the single-particle hybridization through the interatomic potential \cite{Hubsch}. 

In this paper, we study the nonequilibrium spin-$1/2$ single-channel Anderson model with correlated hybridization with the
numerical renormalization-group (NRG) 
approach \cite{Wilson1975,Krishna1,Krishna2}.
We compute the universal zero-bias conductance vs. temperature curve of the 
Kondo regime, and show that deviations from universality can be induced by 
correlated hybridization even when the magnetic susceptibility appears 
universal.  
We find, as predicted earlier \cite{Hubsch} that the spin polarization of the currents can be modestly enhanced by correlated hybridization in some parameter regimes, though not clearly enough to be useful in spintronics applications. We also find that the particle-hole symmetry breaking associated with correlated hybridization can induce a dependence of the Kondo scale upon gate voltage similar to what is observed in experiment \cite{Liang2002,Natelson}.

\begin{figure}
\begin{center}
  $\phantom{a}$\\[-1ex]
  \rotatebox{0}{\scalebox{0.50}{\includegraphics*[0,0][315,225]{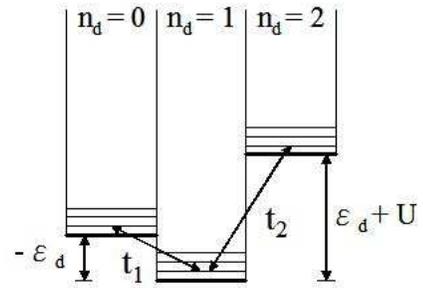}}
}
\end{center}
\caption{
  Schematic for the spin-$1/2$ Anderson model with correlated 
(occupancy dependent) hybridization. Here $t_{1}$ describes transfer of an 
electron into an empty impurity orbital from the leads, and $t_{2}$ describes 
transfer of an electron into 
a singly occupied impurity orbital from the leads.   
}
\label{Fig1}
\end{figure}

To describe the couplings of electrons in the single-molecule transistor, 
we assume the leads are identical and consider
the Hamiltonian for the spin-$1/2$ Anderson impurity model with correlated 
hybridization, as shown schematically in Fig. 1,
\begin{eqnarray}
  \label{H}
\mathcal{H} &=& \sum_{\bf{k},\sigma} \epsilon_{\bf{k}} c_{\bf{k},\sigma}^{\dagger} c_{\bf{k},\sigma} +
      \sum_{\sigma}\epsilon_{d} d_{\sigma}^{\dagger} d_{\sigma} + 
      U n_{d\uparrow} n_{d\downarrow} \\
  && + 
  t_{1} \sum_{\sigma}\left( 1-n_{d,-\sigma}\right)
  \left( d_{\sigma}^{\dagger} c_{\sigma 0} + c_{\sigma 0}^{\dagger} 
  d_{\sigma} \right) \nonumber \\
  && + 
  t_{2} \sum_{\sigma} n_{d,-\sigma}
  \left( d_{\sigma}^{\dagger} c_{\sigma 0} + c_{\sigma 0}^{\dagger} 
  d_{\sigma} \right),
  \nonumber
\end{eqnarray}
where a conduction electron with momentum $\bf{k}$ and spin $\sigma$ 
corresponding to an energy eigenvalue $\epsilon_{\bf{k}}$ is created
by $c_{\bf{k},\sigma}^{\dagger}$, and $c_{\sigma 0}^{\dagger}$ creates a 
conduction electron with spin $\sigma$ at the impurity site
\begin{eqnarray}
\label{c0}
c_{\sigma 0}^{\dagger} = \frac{1}{\sqrt{N}} \sum_{\bf{k}} c_{\bf{k},\sigma}^{\dagger}.
\end{eqnarray}
The operator $d_{\sigma}^{\dagger}$ creates an electron with spin $\sigma$ 
and energy $\epsilon_{d}$ at the d level of the impurity, while $U$ is 
the Coulomb interaction, and $n_{d,\sigma}=d_{\sigma}^{\dagger}d_{\sigma}$ is 
the occupancy of d orbital with spin $\sigma$.

To solve the Hamiltonian, the numerical renormalization group method provides
an excellent approach by mapping the Hamiltonian onto a tight-binding 
Hamiltonian describing a semi-infinite chain with
the impurity at the end. By using the logarithmic discretization and 
diagonalizing the chain Hamiltonian iteratively while keeping the lowest 
eigenstates, one can calculate the thermodynamical and 
dynamical properties with good precision. As a result of the logarithmic
discretization, the hopping along the chain decreases exponentially; the
hopping amplitude between site $n$ and site $n+1$ is proportional to 
$\Lambda^{-n/2}$, where $\Lambda$ is the discretization parameter.

In the quest for new fixed points of the model we investigated the NRG level
flow as well as the temperature-dependent static magnetic susceptibility.
In our calculation, we set
the discretization parameter $\Lambda = 1.65$, and keep the lowest $1500$ 
states in each iteration. The bandwidth $D$ is $50$, and the Coulomb repulsion
$U$ is set to $10$.  The energy scale is fixed by setting the Fermi-level 
density of states per spin $\rho_{F} = 1/\pi$. 

\begin{figure}
\begin{center}
  $\phantom{a}$\\[-1ex]
  \rotatebox{270}{\scalebox{0.35}{\includegraphics*[60,55][555,750]{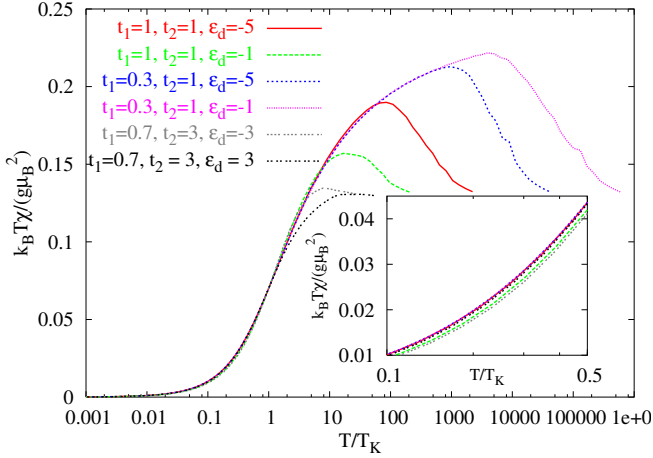}}
}
\end{center}
\caption{
  (color online) Plots of $k_{B}T\chi(T)/(g\mu_B)^2$ vs. $T/T_{K}$ for the Anderson model with correlated
hybridization.  We fix Kondo temperature $T_K$ as the temperature for which 
$k_{B}T\chi/(g\mu_{B})^{2} = 0.07$. The parameters used are
$U=10, D=50$ and as shown in the legend. 
}
\label{Fig2}
\end{figure}

Figure 2 shows the value of $k_{B}T\chi(T)/(g\mu_B)^2$ as a function of 
$T/T_{K}$. 
In the presence of correlated hybridization, the drop of $T\chi$ from the
local-moment regime or mixed-valence regime to the strong-coupling regime looks 
exactly the same as the temperature decreases. 
By analyzing the level flow, we find that there are no novel fixed points 
induced by 
correlated hybridization; all the effects are rolled into a suitable 
renormalization of the Kondo scale. The correlated hybridyzation
may, however, change the steepness of the drop of $T\chi$ if the system goes 
from the free-orbital regime directly to the frozen-impurity regime, 
such as the cases of 
$t_{1}=t_{2}=1$, $\epsilon_{d}=-1$ and $t_{1}=0.7$, $t_{2}=3$, $\epsilon_{d}=-3$,
where there is no Kondo universality.

\begin{figure}
\begin{center}
  $\phantom{a}$\\[-1ex]
  \rotatebox{270}{\scalebox{0.35}{\includegraphics*[60,55][555,750]{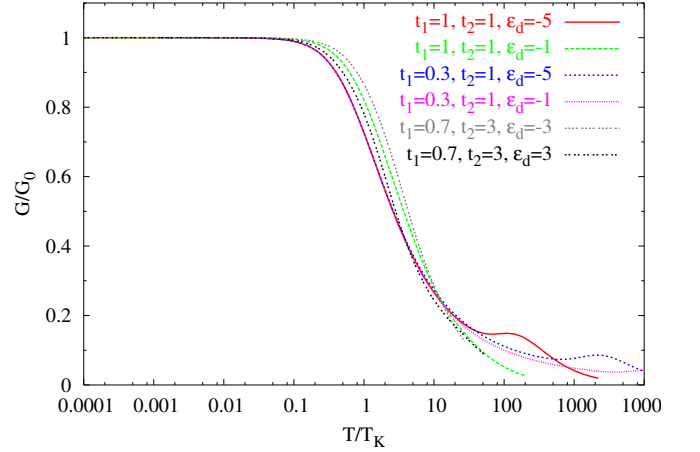}}
}
\end{center}
\caption{
  (color online) Normalized conductance $G(T,V=0)/G_{0}$, where $G_{0}$ is the 
zero-temperature conductance, as a function of $T/T_{K}$.  
}
\label{Fig3}
\end{figure}

The conductance for the single-molecule transistor with symmetric leads 
at zero bias can be expressed as
\begin{eqnarray}
\label{G}
G &=& \frac{e^{2}}{\hbar} \pi \rho_{F} \sum_{\sigma} \int d\omega \left(-\frac{\partial f(\omega)}{\partial \omega}\right)
\rho_{\sigma}(\omega),
\end{eqnarray}
where $\rho_{F}$ is the Fermi-level density of states, $f(\omega)$ is the Fermi distribution function, and 
$\rho_{\sigma}(\omega)$ is the transimission density spectral function. 
We carry out the calculation of transimission density 
$\rho_{\sigma}(\omega)$ in Lehmann representation,
\begin{eqnarray}
\label{rho}
\rho_{\sigma}(\omega) &=& \frac{1}{\pi}\mathrm{Im}\left[ \langle \langle A_{\sigma} | A_{\sigma}^{\dagger}
 \rangle \rangle(\omega - i\delta) \right]
\\
&=& \frac{1}{Z}\sum_{n,m} \left( e^{-\beta E_{n}}+e^{-\beta E_{m}} \right) 
|\langle n|A_{\sigma}|m \rangle |^{2} \\
&& \times
\delta(\omega+E_{n}-E_{m}), \nonumber
\end{eqnarray}
where the operator $A_{\sigma}$ of the Green's function 
$\langle \langle A_{\sigma} | A_{\sigma}^{\dagger} \rangle \rangle(\omega - i\delta)$
is given by
$A_{\sigma}=t_{1}|0\rangle \langle \sigma | - t_{2} |$-$\sigma\rangle 
\langle 2|$ with $|0\rangle$, $|\sigma\rangle$, and $|2\rangle$ being the empty,
 singly, and doubly occupied impurity states, respectively. 
$Z$ is the partition function, and $|n\rangle$,$|m\rangle$ are the
eigenstates of the Hamiltonian in Eq.~\eqref{H}. Therefore, the conductance
can be obtained by
\begin{eqnarray}
\label{Gf}
G = \frac{e^{2}}{\hbar}\frac{\beta}{Z} \pi \rho_{F} \sum_{\sigma,n,m} |\langle n|A_{\sigma}|m \rangle |^{2}
\frac{e^{-\beta (E_{n}+E_{m})}}{e^{-\beta E_{m}}+e^{-\beta E_{n}}}.
\end{eqnarray} 
Since $\rho_{\sigma}(\omega)$ only contributes in the Fermi window, 
the high energy excitation do not contribute, and it is good to take into 
account only the states of the last iteration of the NRG calculation.
The influence of correlated hybridization on the conductance is investigated
by studying the $\epsilon_{d}$ dependence of the conductance. In the molecular
transistor, $\epsilon_{d}$ can be tuned with gate voltage.
As a consequence of correlated hybridization, the symmetry of the 
$0\leftrightarrow 1$ and $1\leftrightarrow 2$ 
valence fluctuations in the conductance is broken, 
with the resonance widths proportional to $t^{2}$
accompanied by the shifts of the resonance positions.
As the temperature decreases from high temperature to the Kondo temperature,
we find the enhancement of the conductance, the broadening of the conductance 
peaks, and the shifts in their positions towards each other, which are 
characteristics of the Kondo effect. In the limit of 
$T\rightarrow0$, the conductance per spin
calculated by the NRG method is found to satisfy the unitary limit, 
$G_{\sigma}=\frac{e^{2}}{h} \mathrm{sin}^{2}\delta_{\sigma}$,
and the peak of total conductance is always found 
$G_{\mathrm{max}}=2e^{2}/h$ to within $\Lambda$-dependent systematic corrections
for different magnitudes of correlated hybridization.   
In the other words, the phase shift analysis using 
Friedel sum rule \cite{Hewson1993} holds with 
correlated hybridization; the Fermi energy phase shift for an electron of spin
$\sigma$ scattering off the impurity site can be obtained
by $\delta_{\sigma}=\pi \langle n_{d,\sigma}\rangle$.

It has been believed that the conductance normalized to its zero-temperature
value is universal in the Kondo regime \cite{Goldhaber1998,Costi1994} as well. 
Figure 3 shows
$G/G_{0}$, where $G_{0}$ is the zero-temperature conductance, as a function of 
$T/T_{K}$. 
We find that for parameters within the range of Kondo regime, 
such as $t_{1} = 0.3$, $t_{2} = 1$, $-5 < \epsilon_{d} < -1$, the curves
fall on top of those without correlated hybridization, showing 
the Kondo universality. However, we notice that 
in the case of $t_{1} = 0.7$ and $t_{2} = 3$, which has a magnitude of 
$|\sigma\rangle \rightarrow |2\rangle$ 
transition comparable to $U$, the conductance curves are obviously off the 
universality, while there is only slight variation in the $T{\chi}$ curves, 
as shown in Fig. 2.
We also note that a perturbative treatment of correlated hybridization in 
Anderson Hamiltonian \cite{Meir2002} was used to explain the anomalous 
conductance plateau around $0.7(2e^{2}/h)$ observed in experiments on 
quantum point contacts. We do not see the $0.7$ anomaly in the conductance 
calculated by this nonperturbative approach. 
In addition, there is no indication of the enhancement of the conductance 
in our calculation induced by local pairing \cite{Guinea2003} associated with 
correlated hybridization.

\begin{figure}
\begin{center}
  $\phantom{a}$\\[-1ex]
  \rotatebox{270}{\scalebox{0.35}{\includegraphics*[60,55][555,755]{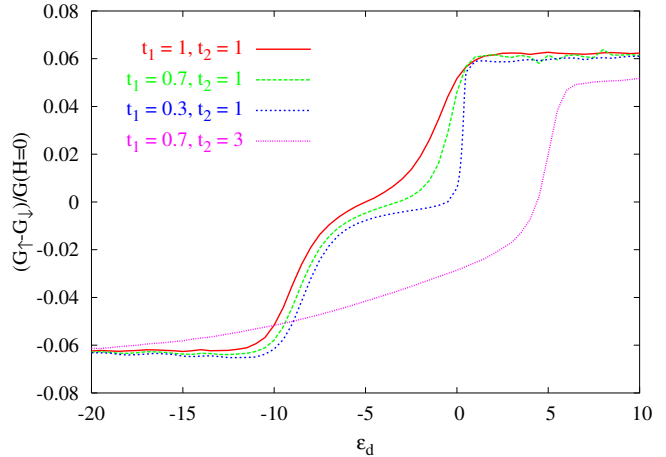}}
}
\end{center}
\caption{
  (color online) The ratio of spin conductance at finite field, 
$H = 0.1k_{B}T_{K}/(g\mu_{B})$, to 
total conductance at zero field in the zero-temperature limit 
as a function of single-particle energy.
}
\label{Fig4}
\end{figure}

Correlated hybridization has been proposed using mean-field treatment 
as a potential means to spin polarization of
currents through transition-metal based molecular transistors in modest
magenetic fields \cite{Hubsch}. However, the effect on spin polarization in the
low-temperature limit is still not clear.
Here, we study the ratio of spin conductance at small magnetic field to 
total conductance at zero field in the zero-temperature limit, 
$\frac{(G_{\uparrow}-G_{\downarrow})}{G(H=0)}$, 
as a function of single-particle energy.
The magnetic field is set to $H=0.1k_{B}T_{K}/(g\mu_{B})$ 
for each set of parameters, as shown in Fig. 4. 
Without correlated hybridization, i.e. $t_{1}=t_{2}=1$, the ratio is zero 
at $\epsilon_{d}=-5$, where there is particle-hole symmetry, and
the zero-field phase shift $\delta_{0}$ = $\pi/2$. 
The ratio $\frac{(G_{\uparrow}-G_{\downarrow})}{G(H=0)}$ increases to about $0.06$ and saturates in the empty and doubly occupied regime. 
When $t_{1}$ is reduced, 
the position of $\delta_{0}=\pi/2$ is shifted towards $\epsilon_{d}=0$ 
because the larger hybridization between singly and doubly occupied impurity 
states renormalizes their energy levels down relative to the level of empty 
impurity state. The spin polarization of currents can be enhanced 
depending upon the value of $\epsilon_{d}$. When $t_{1}=0.7$ and
$t_{2}=3$, there is noticeable enhancement in the spin polarization 
in the regime between $\epsilon_{d}=-3$ and $\epsilon_{d}=-8$. Interestingly, 
the saturated value of the ratio of spin conductance to 
zero-field total conductance remains the same 
for different magnitudes of hybridization, and it seems to have 
universality.

\begin{figure}
\begin{center}
  $\phantom{a}$\\[-1ex]
  \rotatebox{270}{\scalebox{0.35}{\includegraphics*[60,55][555,755]{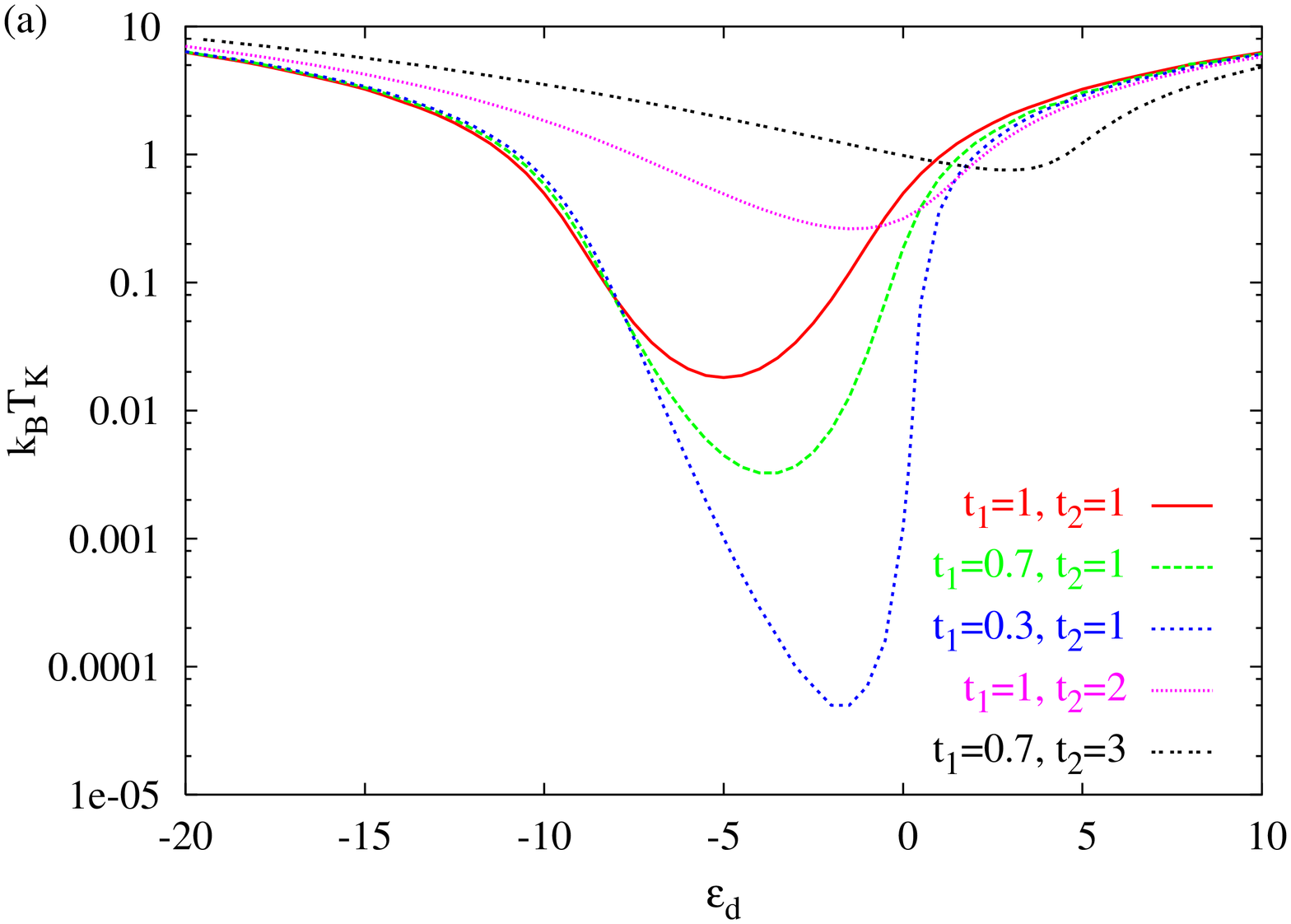}}
}
\end{center}
\begin{center}
  $\phantom{a}$\\[-1ex]
  \rotatebox{270}{\scalebox{0.35}{\includegraphics*[60,55][555,755]{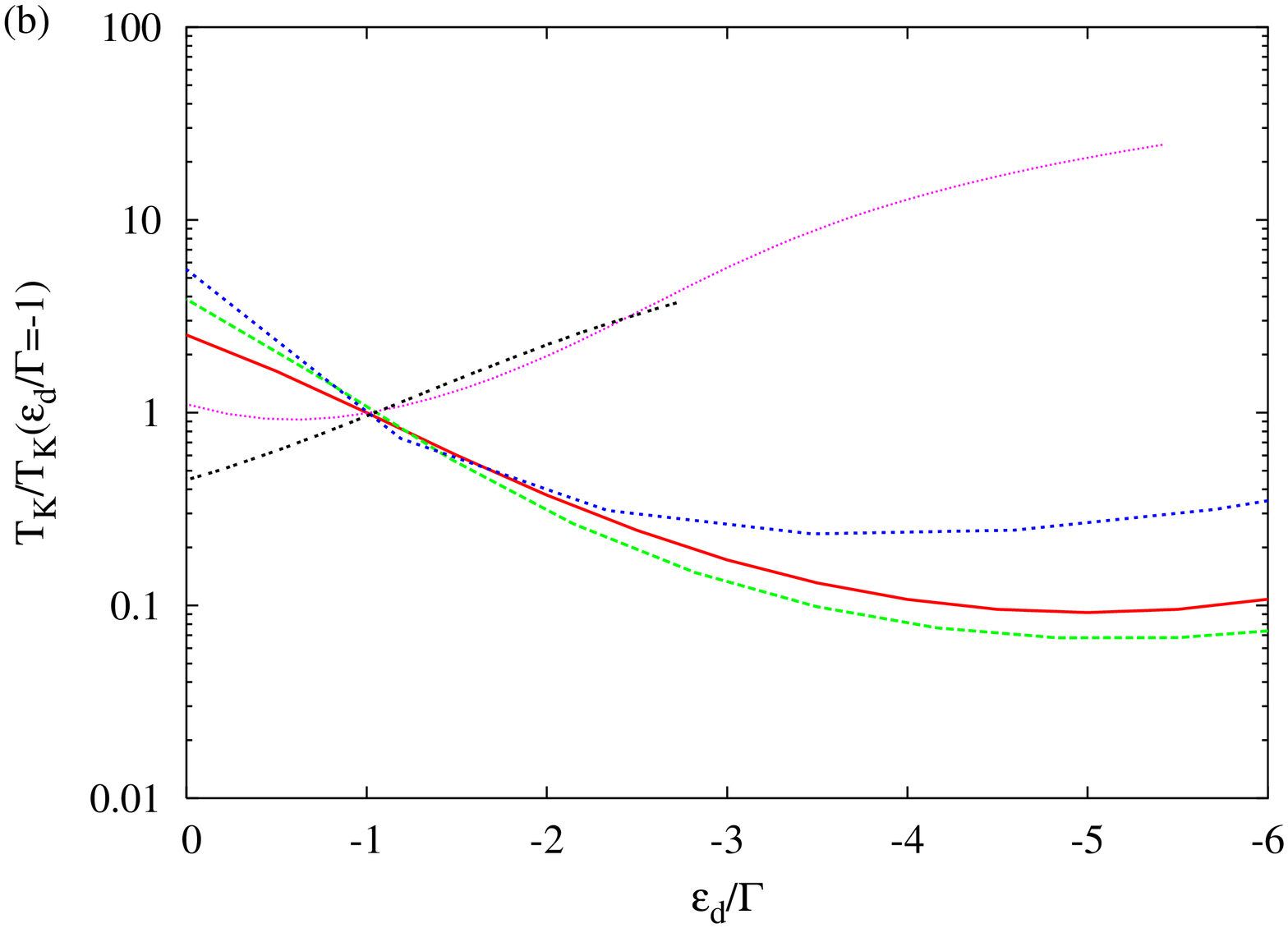}}
}
\end{center}
\caption{
(color online) (a) Kondo temperature as a function of single-particle energy. (b) $T_{K}$ 
(normalized) as a function of $\epsilon_{d}$ normalized by the average
hybridization $\Gamma$. The legend is the same as that in (a).
}
\label{Fig5}
\end{figure}

One of the most interesting results with correlated hybridization is how it 
affects the $\epsilon_{d}$ dependence of the Kondo scale.
Generally without correlated hybridization, as shown in Fig. 5(a), 
the Kondo temperature decreases
when $\epsilon_{d}$ moves away from the charge degeneracy points, 
$\epsilon_{d}=0$ and $\epsilon_{d}=-10$, towards the particle-hole symmetry
point, $\epsilon_{d}=-5$. In the limit of $\Gamma \ll U$, the Kondo 
temperature has been
well described by scaling theory \cite{Haldane1978}, $T_{K}\simeq 1/2 
\sqrt{\Gamma U}$exp$\left[\pi \epsilon_{d}(\epsilon_{d}+U)/(\Gamma U)\right]$, in the Kondo regime.
In the presence of correlated
hybridization, the position of the minimum of Kondo temperature shifts away
from $\epsilon_{d}=-5$. The shift is especially significant in the case of 
$t_{1}=0.7$ and $t_{2}=3$, where there is less $\epsilon_{d}$ dependence
of Kondo temperature.
The particle-hole symmetry breaking associated with correlated hybridization 
may explain the anomalous results in 
a recent Letter \cite{Natelson} that the Kondo 
temperature is less dependent on gate voltage and even increases
when moving away from the supposed charge degeneracy point, $\epsilon_{d} = 0$,
in single-molecule transistors based upon transition-metal complexes. 
In fact, the true charge degeneracy point is shifted away due to 
the renormalization of the energy levels of the impurity states in the presence of correlated hybridization.
Figure 5(b) shows $T_{K}/T_{K}(\epsilon_{d}/\Gamma$=$-1)$ as a function 
of
$\epsilon_{d}/ \Gamma$, where 
$\Gamma=\pi \rho \left[t_{1}(1-\langle n_{d,\sigma}\rangle)+t_{2}\langle 
n_{d,\sigma}\rangle\right]^2$ 
is the average hybridization which has occupancy dependency.
How the Kondo scale changes with $\epsilon_{d}/\Gamma$ depends upon the
magnitude of correlated hybridization.
We see that in the mixed-valence regime
($-1 < \epsilon_{d}/ \Gamma < 0$) the Kondo temperature decreases by about six folds
in the case of $t_{1} = 0.3$, $t_{2} = 1$ when $\epsilon_{d}/ \Gamma$ 
moves away from 0. The decrease of $T_{K}$ becomes
less dramatic when $\epsilon_{d}/ \Gamma$ goes into the Kondo regime 
($\epsilon_{d}/ \Gamma \ll -1$), and the
dependence becomes much weaker when $\epsilon_{d}/ \Gamma < -2$.
The curve seems to capture the feature of the weak gate-voltage
dependence of the Kondo temperature in the single-molecule transistor 
found by Liang {\it et al.} \cite{Liang2002} We note that the orbital degeneracy
which plays a role in Ref. \cite{Liang2002} is absent in our model, so
our calculation of Kondo temperature would have more dependency 
on $\epsilon_{d}/\Gamma$.

The present study shows that the correlated hybridization using spin-$1/2$
single-channel Anderson model can induce deviations in
the normalized zero-bias conductance from universality, and it can modestly 
enhance the spin polarization of currents in small magnetic field. 
The particle-hole symmetry breaking associated with correlated hybridization 
provides an explanation for the weak gate-voltage dependence of the Kondo scale similar to what has been observed in transistion-metal based molecular 
transistors. Further investigations of correlated
hybridization in the degenerate Anderson model are expected to provide
detailed insight into electron transport through a more realistic
single-molecule transistor with orbital degeneracy.

This work was supported by the NSF International Institute for Complex Adaptive Matter (NSF Grant No. DMR-0645461) for J.C. Lin, by the U.S. Department of Energy office of Basic Energy Sciences, Division of Materials Research (J.C.L., D.L.C.) and by the NIC, Forschungszentrum J\"ulich, under project No. HHB000 and 
DFG funding under project AN 275/5-1 (F.B.A.).

\end{document}